\documentclass[pre,aps,showpacs,floatfix,nofootinbib,twocolumn,reprint]{revtex4-1}
\usepackage{epsfig}
\usepackage{amssymb}
\usepackage{amsmath}
\usepackage{amsfonts}
\usepackage{color,graphicx}
\usepackage{bm}

\begin{document}

\title{Nontrivial maturation metastate-average state in a one-dimensional long-range Ising spin glass: above and below the upper critical range}
\author{S.~Jensen$^{1}$, N.~Read$^{2,3}$, and A. P.~Young$^{4}$}
\affiliation{$^{1}$Department of Physics, University of Illinois at Urbana-Champaign, Urbana, Illinois 61801, USA\\
$^{2}$Department of Physics, Yale University, P.O. Box 208120, New Haven, Connecticut 06520-8120, USA\\
$^{3}$Department of Applied Physics, Yale University, P.O. Box 208284, New Haven, Connecticut 06520-8284, USA\\
$^{4}$Physics Department, University of California Santa Cruz, California 95064, USA}

\date{August 16, 2021}

\begin{abstract}
Understanding the low-temperature pure state structure of spin glasses remains an open problem in the field of statistical 
mechanics of disordered systems. Here we study Monte Carlo \emph{dynamics}, performing simulations of the growth of correlations following 
a quench from infinite temperature to a temperature well below the spin-glass transition temperature $T_c$  for a 
one-dimensional Ising spin glass model with diluted long-range interactions. In this model, the probability $P_{ij}$
that an edge $\{i,j\}$ has nonvanishing interaction falls as a power-law with chord distance, $P_{ij}\propto1/R_{ij}^{2\sigma}$,
and we study a range of values of $\sigma$ with $1/2<\sigma<1$. We consider a correlation function $C_{4}(r,t)$. A
dynamic correlation length that shows power-law growth with time $\xi(t)\propto t^{1/z}$ can be identified 
in the data and, for large time $t$, $C_{4}(r,t)$ decays as a power law $r^{-\alpha_d}$ with distance $r$ when 
$r\ll \xi(t)$. The calculation can be interpreted in terms of the {\em maturation} 
metastate averaged Gibbs state, or MMAS, and the decay exponent $\alpha_d$ differentiates between a trivial MMAS 
($\alpha_d=0$), as expected in the droplet picture of spin glasses, and a nontrivial MMAS ($\alpha_d\ne 0$), as in the 
replica-symmetry-breaking (RSB) or 
chaotic pairs pictures. We find nonzero $\alpha_d$ \emph{even} in the regime $\sigma >2/3$ which corresponds to 
short-range systems below six dimensions. For $\sigma < 2/3$, the decay exponent $\alpha_d$ follows the RSB prediction 
for the decay exponent $\alpha_s = 3 - 4 \sigma$ of the {\em static} metastate, consistent with a conjectured 
statics-dynamics relation, while it approaches $\alpha_d=1-\sigma$ in the regime $2/3<\sigma<1$; however, 
it deviates from both lines in the vicinity of $\sigma=2/3$.
\end{abstract}

\maketitle

\section{Introduction}
\label{intro}

The low-temperature equilibrium pure-state structure of classical Ising spin glasses has been debated for many years, 
and is still not well understood. The Ising spin glass models~\cite{Edwards1975} are defined with discrete two-state 
spin variables interacting on a $d$-dimensional hypercubic lattice ($s_i =\pm 1$ for a spin at lattice site $\bold{r}_i$ 
with lattice spacing 1) with the Hamiltonian 
\begin{equation}\label{eq:Hamiltonian}
H(S)=-\sum_{\{i, j\}}J_{ij}s_{i}s_{j}\;,
\end{equation}
where the bonds $J_{ij}=J_{ji}$ for each undirected edge (unordered pair) $\{i,j\}$ are independent random variables 
which form a collection $\mathcal{J}\equiv\left (J_{ij}\right )_{\{ij\}}$. A spin configuration 
is denoted $S\equiv \left(s_{i}\right)_i$. The edges connect all distinct sites $i\ne j$, $\bold{r}_{i}\in \bold{Z}^d$ 
where $d$ is the dimension of space
(for a finite lattice $\Lambda$, $\bold{r}_{i}\in \Lambda$ with $\Lambda\subset \bold{Z}^d$ with some chosen 
boundary conditions) of the graph with vertices $i$ and edges $\{i,j\}$. The probability distribution over the bonds 
$\mathcal{J}$ is denoted $\nu(\mathcal{J})$ and the disorder average over the 
bonds distribution is then denoted $\left [ \cdot \cdot \cdot \right ]_{\nu(\mathcal{J})}$. 

For an \emph{infinite} system there can be a phase transition which necessarily presents ergodicity breaking, in which 
the configuration space can be divided into disjoint regions, such that under time evolution the system remains forever 
in one region, but explores all of it (the dynamics restricted to one region is ergodic). An example of this is the 
ferromagnetic Ising model defined as for Eq.~\eqref{eq:Hamiltonian} but with constant couplings $J_{ij}=J>0$ for 
edges connecting nearest neighbors ($|\bold{r}_{i}-\bold{r}_{j}| =1$ with the Euclidean metric). In thermal equilibrium,
below some nonzero critical temperature $T_c$ (for spatial dimension $d \geq 2$) there is spontaneous symmetry breaking 
with at least two ``pure'' (or ordered) states, denoted $\Gamma_{\uparrow}$ ($\Gamma_{\downarrow}$) for the 
``up'' (``down'') state. Under time evolution of the system, there is zero probability that an initial configuration $S$ 
drawn from the up pure state will be found in the down pure state, or in any pure state other than the up state (and 
similarly for initial $S$ in the down state). [Here we restrict the discussion to boundary conditions imposed as  
$s_{i}=+1$ for $s_{i}\in \partial \Lambda$ for the up state and $s_{i}=-1$ for $s_{i}\in \partial \Lambda$ for 
the down state (with $\Lambda \rightarrow \infty$).] These two states are the only translationally-invariant pure Gibbs 
states for the ferromagnetic Ising model below the phase transition temperature $T_c$ in any dimension $d\geq 2$. 
Generally, ergodic states for the dynamics are the same as equilibrium pure states.

Unlike for the ferromagnetic Ising model, for the general spin glass model of Eq.~\eqref{eq:Hamiltonian}, the number 
and nature of pure states in the low-temperature phase is not clear. This has been debated extensively for short-range 
spin glasses including the Edwards-Anderson (EA) model~\cite{Edwards1975} in low dimension $d$, following the 
early works of Refs.~\cite{Parisi1979, Parisi1980, Parisi1983} which developed the theory of so-called replica-symmetry 
breaking (RSB) as a mean-field theory in infinite-range models, and of 
Refs.~\cite{Bray1984, McMillan1984, Bray1985, Fisher1986, Fisher1988} where the scaling-droplet (SD) theory of finite-range 
models was developed. In the EA model, only the nearest-neighbor bonds are nonzero, and they are identically-distributed 
Gaussians with vanishing mean and variance unity, 
\begin{equation}
\left [ J_{ij} \right ]_{\nu(\mathcal{J})} = 0, \textrm{        }\left [ J_{ij}^2\right ]_{\nu(\mathcal{J})}=1\;.
\end{equation}
Determining the pure state structure for this model with analytical work is challenging as there is no controlled approach known 
for low dimension $d$. 

We will consider a dynamical correlation function; in a moment we will further explain its relation to equilibrium properties.
The function is defined by 
\begin{equation}\label{eq:C4_equation}
C_{4}(\bold{r}_{i}-\bold{r}_{j},t)\equiv \left [ \left[\langle s_{i}(t)s_{j}(t)\rangle_{S|S_0}\right]_{\eta(S_0)}^{2}
\right ]_{\nu(\mathcal{J})},
\end{equation}
where (i) $\langle \cdot\cdot\cdot\rangle_{S|S_0}$ denotes an average over trajectories $\{S(t'):0< t'\leq t, S(0)=S_0\}$ 
of $S$ in time $t$, with initial value $S_0$ at $t=0$, under some stochastic dynamics that has the Gibbs distribution 
at temperature $T$ as its stationary state (which is unique in a finite size system; in practice, we will use Monte Carlo 
evolution), and (ii) $[\cdots]_{\eta(S_0)}$ is expectation over a distribution $\eta(S_0)$ of 
initial configurations $S_0$, which is an infinite temperature state (i.e.\ the uniform distribution on 
spin configurations).  This corresponds 
to dynamic evolution of the correlation function following an instantaneous quench from $T=\infty$ to a final temperature 
$T$, and we will choose $T$ to be well below the equilibrium transition temperature $ T_c$. Such a correlation function has 
been studied previously, in Refs.~\cite{Marinari1996, Belletti2009, Manssen2015, Manssen2015-2,Baity-Jesi2018} for the 
EA model, in 
Ref.~\cite{Wittmann2016} for a one-dimensional diluted long-range interacting model (which we will describe later), and
for other models in Ref.\ \cite{White2006}. In each case it was expected that
$C_{4}(\bold{r}_{i}-\bold{r}_{j},t)$ would follow a scaling ansatz
\begin{equation}\label{eq:decay}
C_{4}(\bold{r}_{i}-\bold{r}_{j},t)=\frac{1}{r_{ij}^{\alpha_d}}f\left (\frac{r_{ij}}{\xi(t)} \right )\;,
\end{equation}
where $r_{ij}\equiv |\bold{r}_{i}-\bold{r}_{j}|$, $f(x)$ is a scaling function [$f(x)$ tends to a constant as $x\to0$], 
$\xi(t)$ is a dynamical correlation length, and $\alpha_d$ is the dynamic spatial decay exponent. The correlation length 
was expected to behave as a power law with time $\xi(t)\propto t^{1/z}$ for large time, where $z<\infty$ is a dynamical 
exponent. For $r_{ij}\ll\xi(t)$ this gives power-law decay $C_{4}(\bold{r}_{i}-\bold{r}_{j},t)\propto 1/r_{ij}^{\alpha_d}$. 
One would expect that $\alpha_d$ is independent of $T$ for $0<T<T_c$, while $z=z(T)$ has been found to depend on $T$ 
\cite{Wittmann2016}. The ansatz was found to hold numerically (for the times and system 
sizes studied) with varying degrees of accuracy in Refs.~\cite{Marinari1996, Belletti2009, Manssen2015, 
Manssen2015-2,Baity-Jesi2018,Wittmann2016}.

If the power-law form indeed holds asymptotically in an infinite size system, with $\alpha_d>0$, it implies that
as $t\to\infty$ the system reaches a statistical state described by expectations of the form $\lim_{t\to\infty}[\langle \cdots\rangle
_{S|S_0}]_{\nu(S_0)}$ that are independent of $t$ in the limit, with decay of equal-time correlations 
$\lim_{t\to\infty}[\langle s_i(t)s_j(t)\rangle_{S|S_0}]_{\eta(S_0)}$ to zero with distance, at least in the sense of the 
disorder average of the square. This differs dramatically from what should occur if the state in the long-time limit (which 
we assume is stationary, though it is not obvious this must hold; we return to this later) is what we will call a 
trivial Gibbs state, that is one that (here again for $T<T_c$) is the equal-weight mixture of two pure states that are related 
by spin-flip symmetry, as the SD picture \cite{Bray1984, McMillan1984, Bray1985,Fisher1986,Fisher1988} 
assumes is the case in equilibrium at zero magnetic field. In the latter case the correlation function $\lim_{t\to\infty}C_4
(\bold{r}_{i}-\bold{r}_{j},t)$ would go to a positive constant as $r_{ij}\to\infty$, which means $\alpha_d=0$. 
Thus $\alpha_d>0$ should imply that there are 
(infinitely) many pure states, which are accessed by the protocol that defines $C_4$. We  will view the state obtained 
at long times, averages in which are $\lim_{t\to\infty}[\langle\cdots\rangle_{S|S_0}]_{\eta(S_0)}$, as the 
maturation-metastate--average state or MMAS (see Ref.\ \cite{White2006}; the term metastate 
is used by analogy with the metastate in statics~\cite{Aizenman1990,Newman1996-2, Newman1997,Newman2003}). 
We further explain some of this in the following Section; the value of $\alpha_d$ is quantitative information about the MMAS, and 
$\alpha_d>0$ means the MMAS contains many pure states. 

In Ref.~\cite{Wittmann2016}, a one-dimensional diluted long-range interacting model was considered,
which makes possible the use of very large (i.e.\ long) systems in which to consider the correlations. This model is a diluted, 
non-Gaussian variant (described below) of a well-known one-dimensional model~\cite{Kotliar1983} that has independent 
Gaussian distributions for the bonds $\mathcal J$; in both models, the bonds $J_{ij}$ for each pair $\{i,j\}$ have mean zero 
and variance
\begin{equation}
\left [ J_{ij}^2\right ]_{\nu(\mathcal{J})}\propto \frac{1}{r_{ij}^{2\sigma}}\;
\label{kotdef}
\end{equation}
as $r_{ij}\to\infty$.
These models have a transition with $T_c>0$ for $1/2<\sigma<1$, and are sometimes considered as a proxy for a 
short-range interacting model with $\sigma$ playing the role of $d$.
Ref.~\cite{Wittmann2016} considered a single value $\sigma=0.625$ for which a suggested value of $\alpha_d$ was available 
(we discuss this in the following Section), and obtained excellent agreement with that value. 

In this paper we extend the study of 
Ref.~\cite{Wittmann2016} for the diluted long-range one-dimensional model to a wide range of $\sigma<1$. We find 
a nontrivial metastate ($\alpha_d>0$) for all such interactions $\sigma$. We also find support, 
in agreement with Ref.~\cite{Wittmann2016}, for a conjectured statics-dynamics relation involving $\alpha_d$ in 
the region $\sigma<2/3$, and some support for an empirical interpolating form in the complementary regime 
$2/3\leq\sigma<1$.

Some additional background material is presented in Section \ref{sec:back}, while the details and results of simulations are 
given in Section \ref{sec:results}. Appendix \ref{app:best_fit} discusses the methods used to obtain the best fits in the scaling 
collapse plots.

\section{Background} 
\label{sec:back}

Here we will briefly review and explain a number of concepts to which we will refer, including states, Gibbs states, pure states,
and metastates and metastate-average states, both static and dynamic, along with some of their properties. In general, 
in this Section, systems are assumed to be of infinite size unless stated to be finite. 

\subsection{Equilibrium (Gibbs) states and pure states}

First, by a state of an Ising spin system we always mean a probability distribution on spin configurations $S$. In finite-range
spin systems, a state of thermal equilibrium is usually assumed to be a Gibbs state. We fix a choice of $\mathcal J$ 
throughout the discussion of Gibbs and pure states. In a {\em finite}-size system, 
a Gibbs state $\Gamma_{\mathcal J}$ for a given Hamiltonian $H$ as in Eq.\ (\ref{eq:Hamiltonian}) and temperature $T$ 
is defined by
\begin{equation}
\Gamma_{\mathcal J}(S)=e^{-H(S)/T}/\sum_S e^{-H(S)/T}.
\label{finsizgibbs}
\end{equation}
In an infinite system, this formula cannot be used directly. Instead, 
a Gibbs state $\Gamma_{\mathcal J}$ is defined by the Dobrushin-Lanford-Ruelle (DLR) conditions \cite{georgii_book} 
which say that, for any finite subset $\Lambda$, the conditional probability distribution for the spins 
$S|_\Lambda=(s_i)_{i\in\Lambda}$ at sites in $\Lambda$, conditioned on the remaining spins $S|_{\Lambda^c}$ in the 
complement $\Lambda^c$ of $\Lambda$, is 
\begin{equation}
\Gamma_{\mathcal J}(S|_\Lambda\mid S|_{\Lambda^c})=e^{-H_\Lambda'(S)/T}/\sum_{S|_\Lambda} 
e^{-H'_\Lambda(S)/T},
\label{dlrcond}
\end{equation}
where $H_\Lambda'(S)$ is the sum of only the terms $-J_{ij}s_i s_j$ in which at least one of $i$, $j$ is in $\Lambda$. 
As these conditions never specify what happens at infinity, there may be many distinct Gibbs states 
that satisfy the same conditions, and the appearance of such non-uniqueness at low temperature describes one possible way 
in which a phase transition can occur.

From the definition, a convex combination (or ``mixture'') of Gibbs states is again a Gibbs state.
A pure (or extremal) Gibbs state is one that cannot be expressed as a convex 
combination of other Gibbs states; the pure states form a subset of the set of all Gibbs states. Any Gibbs state 
$\Gamma_{\mathcal J}$, say, can be expressed (or decomposed) uniquely as a convex combination of pure Gibbs states 
in the form \cite{georgii_book}
\begin{equation}
\Gamma_{\mathcal J}=\int d\varepsilon \, w_{{\mathcal J}\Gamma_{\mathcal J}}(\varepsilon)\Gamma_{{\mathcal J}
\varepsilon},
\end{equation}
where $w_{{\mathcal J}\Gamma_{\mathcal J}}(\varepsilon)$ is a probability distribution on pure states 
$\Gamma_{{\mathcal J}\varepsilon}$ labeled by $\varepsilon$. $w_{{\mathcal J}\Gamma_{\mathcal J}}(\varepsilon)$,
which depends on both $\mathcal J$ and the chosen Gibbs state $\Gamma_{\mathcal J}$, is called the weight of the
decomposition; we have written it as an integral for generality, but the decomposition might reduce to a sum of a 
countable number of terms. 

\subsection{Equilibrium (static) metastate}

A static, or equilibrium, metastate, denoted $\kappa_{\mathcal{J}}$, is a probability distribution on Gibbs states 
$\Gamma_{\mathcal{J}}$ in infinite size that is obtained by taking a limit of finite-size systems. There are 
a couple of different constructions of an equilibrium metastate. For a system of finite size $L$, we will write 
$\langle\cdots\rangle$ for an expectation in the unique equilibrium state, which depends on the disorder (the bonds)
$\mathcal J$. In the Aizenman-Wehr (AW) metastate \cite{Aizenman1990}, the metastate average of a quantity is defined
by first taking the expectation with respect to $\nu$ but only for the bonds $J_{ij}$ with both $i$, $j$ a distance 
greater than $M<L$ from the origin (i.e.\ in the outer region); denote that by $[\cdots ]_>$, and the average over the remaining 
bonds (``in the inner region'') by $[\cdots]_<$. On taking the limits $L\to\infty$, then $M\to\infty$, these are denoted 
by the metastate average $\left [ \cdot\cdot\cdot\right ]_{\kappa_{\mathcal{J}}}$ (for given $\mathcal J$ in the inner region), 
and by $[\cdots]_{\nu(\mathcal{J})}$, respectively; the inner region is now infinite in size, so we can use the same notation 
$\mathcal J$ for the disorder there, and $\nu$ for its distribution. As the equilibrium state depends on the disorder in the outer 
region, in the $L\to\infty$, $M\to\infty$ limit it becomes a Gibbs state $\Gamma_{\mathcal J}$ that, even in a finite region 
near the origin, may retain some dependence on the disorder in the outer region far away (as well as on $\mathcal J$), 
and if it does then the metastate is nontrivial (i.e.\ it is supported on more than one Gibbs state). We write 
$\langle\cdots\rangle_{\Gamma_{\mathcal J}}$ for the thermal expectation in $\Gamma_{\mathcal J}$. 
The Newman-Stein (NS) metastate \cite{Newman1996-2, Newman1997,Newman2003} is similar, except that the 
average over disorder at distance $>M$ from the origin is replaced by an average over a range of system sizes between 
$M$ and $L$ at given disorder; we will use the same notation for either construction. Both constructions
require some further discussion of the limits (for example, the possible need to take the limit along a subsequence 
of sizes $L$ and $M$), for which see
\cite{Aizenman1990,Newman1996-2, Newman1997,Newman2003}. Finally, it is useful to define the average 
of the Gibbs state $\Gamma_{\mathcal J}$ over the metastate $\kappa_{\mathcal{J}}$, which produces another Gibbs 
state, the metastate-averaged state (MAS), denoted $\rho_{\mathcal{J}}$. That is, a MAS thermal correlation function is 
calculated as
\begin{equation}
\langle\cdot\cdot\cdot\rangle_{\rho_{\mathcal{J}}}\equiv\left [ \langle\cdot\cdot\cdot\rangle_{\Gamma_{\mathcal{J}}}
\right ]_{\kappa_{\mathcal{J}}}.
\label{rhodef}
\end{equation}

\subsection{Long-range models}

As we are concerned in this paper with long-range spin-glass models, rather than the short-range ones that were implicit in 
the discussion so far, it needs to be said before going further that the definition of a Gibbs state
as in Eq.\ (\ref{dlrcond}) breaks down in that case. That is because for typical $\mathcal J$ the sum in $H_{\Lambda}'(S)$ 
does not converge absolutely, and diverges for some $S$, when $\sigma<1$ [see the definition in Eq.\ (\ref{kotdef})]. 
Consequently, pathological states 
exist in the model, and the definition of a Gibbs state should be modified so that only converging sums occur 
\cite{Gandolfi1993}. Within a metastate construction, such pathologies do not occur, and only Gibbs states in the modified 
sense arise \cite{Read2021}. The same issues should also be addressed for the maturation metastate, but this has not 
been carried out so far. For that we will proceed on the assumption that these technical issues do not obstruct what we will 
discuss. (Of course, simulations are performed in finite systems, for which the issue does not arise.)

\subsection{Correlations in the equilibrium MAS}

Ref.~\cite{Read2014} introduced a correlation function in the MAS,
\begin{eqnarray}
C(\bold{r}_{i}-\bold{r}_{j}) & \equiv& \left [ \langle s_{i} s_{j}\rangle_{\rho_{\mathcal{J}}}^2\right ]_{\nu(\mathcal{J})}
\label{C_def}\\ 
&=&\left [\left [ \langle s_{i} s_{j}\rangle_{\Gamma_{\mathcal{J}}}
\right ]_{\kappa_{\mathcal{J}}}^2\right]_{\nu(\mathcal{J})}.
\end{eqnarray}
Note that, by eq.\ (\ref{rhodef}), both the thermal expectation and the metastate average are performed {\em before} 
the square is taken,
which, for example for the AW metastate, differs from the traditional average over all disorder at once (for 
which, see below). The large-distance behavior of this correlation function in the low-temperature phase was predicted 
to be
\begin{equation}
\label{eq:power_law_form}
C(\bold{r}_{i}-\bold{r}_{j})\sim \frac{1}{r_{ij}^{\alpha_s}}\;,
\end{equation}
as $r_{ij}\to\infty$, up to a constant factor, with a decay exponent $\alpha_s \ge 0$. $\alpha_s>0$ implies that there are 
many pure states in the decomposition of the MAS $\rho_{\mathcal J}$, and (presumably) that the metastate is non-trivial, 
as we will explain; as for $\alpha_d$, one expects the value to be the 
same for all $0<T<T_c$. (This form also holds in some of the models in Ref.\ \cite{White2006}.) It was further shown in 
Ref.~\cite{Read2014} that the Landau-Ginzburg field theory of RSB in a finite-range spin 
glass leads to a description with a non-trivial metastate for $T<T_c$, and that
\begin{equation}
\alpha_s=d-4
\end{equation}
for $d>6$ where the calculation can be done. For the one-dimensional power-law models mentioned at the end of the preceding
section, this formula becomes
\begin{equation}
\alpha_s=3-4\sigma
\end{equation}
for $1/2<\sigma\leq 2/3$ \cite{Wittmann2016}. The latter region, to which we may refer as $\sigma$ below the upper critical 
range, is also that in which the critical exponents at $T=T_c$ take their mean-field values (as in a short-range system for 
$d$ above the upper critical dimension, which is $d=6$ for spin glasses), while the region $\sigma>2/3$ is above the 
critical range, and some of the critical exponents for $2/3<\sigma < 1$ differ from their mean-field values. It is natural to 
expect similar phenomena for $\alpha_s$ and $\alpha_d$, even though they are defined for $T<T_c$, because the perturbative 
field theory approach for correlations runs into (so far unresolved) difficulties for $T<T_c$ when $d<6$ or $\sigma>2/3$
that are more severe than those for $T=T_c$ (where the renormalization group allows calculation of the exponents). 

In the SD picture of spin glasses, the metastate is tacitly assumed to be trivial, and
\begin{equation}
\underset{r_{ij}\rightarrow\infty}{\textrm{lim}}C(\bold{r}_{i}-\bold{r}_{j})=q^2\;,
\end{equation}
where $q$ is the order parameter, so $\alpha_s$ is then defined to be zero. The order parameter 
would be defined in general as the limit of the spin-glass correlation function
\begin{equation}
\lim_{r_{ij}\rightarrow\infty}\left [ \left[\langle s_is_j\rangle^2_{\Gamma_{\mathcal{J}}}
\right ]_{\kappa_{\mathcal{J}}}\right]_{\nu(\mathcal{J})}= q^{(2)} \;.
\label{q2}
\end{equation}
Note the crucial difference from the MAS correlation function $C$ in Eq.~\eqref{C_def};
the square in Eq.~\eqref{q2} is taken \textit{before} the metastate average, and for the AW metastate 
$\left[[\cdots]_{\kappa_{\mathcal J}}\right]_{\nu({\mathcal J})}$ corresponds simply to the traditional average 
over all disorder. If the metastate is non-trivial then the left-hand side of 
Eq.~\eqref{q2} (without the $r_{ij}\to\infty$ limit) is different from $C$. In terms of RSB, $q^{(2)}=\int_0^1 q(x)^2\,dx$ 
\cite{Parisi1983}, while in the SD picture $q(x)^2=q^2$ is constant. 
In RSB, $C$ tends to $q(0)^2$ \cite{Read2014}, which is zero in zero magnetic field, and $q(0)^2\leq q^{(2)}$ 
because $q(x)^2$ is an increasing function of $x$. In general, we can {\em define} $q(0)^2$ by 
$q(0)^2=\lim_{r\to\infty}C(r)$, and then $q(0)^2\leq q^{(2)}$ always, but $q(0)^2$ is not necessarily zero
(see below for further discussion of this point). 
Then
\begin{equation}
q(0)^2<q^{(2)}
\label{q0q2}
\end{equation} 
always implies a nontrivial metastate. The SD picture 
of spin glasses is the only scenario with a trivial metastate and trivial Gibbs state. The chaotic pairs picture 
\cite{Newman1996-2, Newman1997,Newman2003} has a nontrivial metastate supported on trivial Gibbs states;
in that case $q(x)^2$ is a constant, larger than $q(0)^2$, for all $x>0$ \cite{Read2014}, and the power-law form 
with $C$ tending to zero is valid in some cases \cite{White2006} there also, though possibly not 
always.  An accurate and reliable calculation of the static MAS correlation function would then partially resolve the debate 
about the low-temperature structure for a spin glass model. Calculating the exponent $\alpha_s$ for low dimension $d$ 
however remains difficult but there has been recent numerical progress with a Monte Carlo study of the EA model in 
Ref.~\cite{Billoire2017}.

\subsection{Maturation MAS (MMAS)}

There is an evident similarity or analogy between the dynamical MMAS defined by expectations $\lim_{t\to\infty}
[\langle\cdots\rangle_{S|S_0}]_{\eta(S_0)}$ and the static MAS defined by $\langle\cdots\rangle_{\rho_{\mathcal J}}$, 
and between their corresponding correlation functions $C_4$ and $C$, respectively. First, if the MMAS exists as a limit, 
it is plausible that it must be a stationary state, and stationary states are believed to be necessarily Gibbs states
(this has been proved in the translation-invariant case; see e.g.\ Ref.\ \cite{liggett_book}). 
For example, consider one picture of the evolution of the state from a given random initial condition $S_0$, and
assume the validity of the SD picture. At long times there will be domains, within each of which the state locally
can be approximated by one of the two pure states, and the domains will be separated by domain 
walls where the state changes to the other pure state; the scale of the domains increases with time as $\xi(t)$. (In the SD 
theory, $\xi(t)$ is expected to diverge as a power of $\ln t$, not as a power of $t$ \cite{Fisher1988b}.) 
The domain walls should be sparse \cite{Fisher1988}, so the probability that one separates a given ${\bf r}_i$ from 
a given ${\bf r}_j$ at time $t$ for given $S_0$ eventually goes to zero as $t\to\infty$. Hence within this picture we 
expect that the $t\to\infty$ limit of the $\eta(S_0)$-average state {\em is} a stationary state, which is the trivial Gibbs 
state. Note however that the state for given $S_0$, for example in any fixed finite region, does not tend to a limit, but 
keeps switching.

Second, the static MAS is an average of the state (correlations) of the spins near the origin with respect to either the disorder 
far away, or the finite system size, and we will show that this average may reveal that there are many pure states of the 
infinite system, even when a single Gibbs state drawn from $\kappa_{\mathcal J}$ only involves a smaller number 
(as in the RSB and chaotic pairs pictures). Similarly, the dynamic MMAS is the long-time limit of the average of the 
equal-time correlations with respect to the 
initial conditions, and this average too may show that there are many pure states; the initial configuration can affect
the state far from the region of interest at long times, somewhat like the distant disorder. The MAS and the MMAS may thus 
be very closely related, or possibly identical (a similar remark appears in Ref.\ 
\cite{Newman1999}). To sharpen the analogy, we denote the MMAS by $\rho_{\mathcal J}^{\rm M}$, and so 
$\langle\cdots\rangle_{\rho_{\mathcal J}^{\rm M}}= \lim_{t\to\infty}[\langle\cdots\rangle_{S|S_0}]_{\eta(S_0)}$.

It is tempting to go further and try to define a maturation metastate \cite{White2006} $\kappa_{\mathcal J}^{\rm M}$, 
a distribution on Gibbs states $\Gamma_{\mathcal J}$, such that $\langle\cdots\rangle_{\rho_{\mathcal J}^{\rm M}}
=\left[\langle\cdots\rangle_{\Gamma_{\mathcal J}}\right]_{\kappa_{\mathcal J}^{\rm M}}$. We are not aware of a formal
treatment of such a construction, but the initial steps (similarly to the equilibrium metastate \cite{Aizenman1990,Newman1997}),  
might be to consider (in infinite size) the joint distribution of the state (not the spins), the bonds, and the initial 
configuration, take the $t\to\infty$ limit (possibly using a subsequence), sum over initial conditions, and then condition 
on the bonds to obtain $\kappa_{\mathcal J}^{\rm M}$. One would then want to know that the states drawn from 
$\kappa_{\mathcal J}^{\rm M}$ are Gibbs states. If so, the analogs of the general statements above
for the static metastate, such as Eq.\  (\ref{q0q2}), would also hold for the maturation metastate.
Variations of this construction can also be considered; for example, the random variables involved in the dynamics up 
to a time $t^*>0$ with $t^*<t$ (in other words, $S(t')$ for $0\leq t'\leq t^*$) can be treated as part of the initial 
conditions along with $S_0$, with only the subsequent evolution producing the state. In these constructions, the 
maturation metastate average $[\cdots ]_{\kappa_{\mathcal J}^{\rm M}}$ of a quantity is the average over the initial 
segment $S(t')$ for $0\leq t'\leq t^*$ (including $S_0$), with suitable limits taken, analogously to the AW static metastate, 
and exactly as stated informally in the preceding paragraph.

Instead of the long-time limit, the literature generally focuses on the state in a finite region at a finite time after 
the quench, and attempts to describe how the limit $t\to\infty$ is approached. In particular, we can ask whether, 
conditioned on $S_0$ and on the dynamical randomness up to time $t^*<t$, the 
subsequent time evolution to $t\to\infty$ (with $t^*\to\infty$ as some function of $t$) produces a {\em pure} state 
(for a more precise discussion, see Ref.\ \cite{Newman1999}); in that case, the behavior described above, in which any 
fixed finite region switches infinitely often from one pure state to another (the phenomenon of  ``local non-equilibration'' 
\cite{Newman1999}), is excluded. If that holds, then a theorem of NS (Theorem 2 in Ref.\ \cite{Newman1999}) shows 
that the number of pure states in the decomposition of $\rho_{\mathcal J}^{\rm M}$ (for which see below also) must be 
uncountable, and it also follows from their result that $\lim_{t\to\infty}C_4(r,t) \to0$ as $r\to\infty$, ruling out 
$\alpha_d=0$. [Stated differently, 
NS's result shows that for the SD picture, the state that evolves from a given $S_0$ and given dynamical randomness up to 
$t^*$ must exhibit local non-equilibration, no matter how $t^*$ diverges as $t$ does.] When the hypothesis holds, the 
corresponding $\kappa_{\mathcal J}^{\rm M}$ becomes a distribution on {\em pure} Gibbs states, but again the 
general statements remain valid.

\subsection{Correlations in the MMAS}

In the remainder of this paper, we will consider only the MMAS, which is simpler to define and study numerically, and which 
has a close relation with the static MAS. We will use the terms trivial or nontrivial for the MMAS at $T<T_c$ in the following way: 
$\alpha_d=0$ is considered the trivial case, 
and occurs if there is a finite or countably infinite number of pure states in the MMAS that each have nonzero weight, 
while $\alpha_d>0$ is considered nontrivial, and implies that (i.e.\ occurs only if) there is an uncountably infinite number 
of pure states involved and no one pure state has nonzero weight. To explain this, first, the term ``weight'' refers to the 
decomposition of the MMAS, which we assume is a Gibbs state, into pure states: 
\begin{equation}
\rho_{\mathcal J}^{\rm M}=\int d\varepsilon \,\mu_{\mathcal J}^{\rm M}(\varepsilon)\, \Gamma_{{\mathcal J}\varepsilon},
\end{equation}
where again $\Gamma_{{\mathcal J}\varepsilon}$ is a pure state for the given $\mathcal J$, and the probability measure 
$\mu_{\mathcal J}^{\rm M}(\varepsilon)$ on the pure states $\varepsilon$ could be continuous, or could consist solely 
of $\delta$-functions so that the integral reduces to a simple sum of weights on a countable collection of pure states, or 
could be a combination of both.  (There is a completely parallel analysis for the static MAS $\rho_{\mathcal J}$, with 
corresponding weight $\mu_{\mathcal J}$.) Next, as 
$\lim_{t\to\infty}C_4$ is supposed to tend to a limit as $r\to\infty$, 
it will make no difference to the value of that limit if we average both ${\bf r}_i$ and ${\bf r}_j$ over a hypercubic box 
$\Lambda_W$ of $W^d$ sites, and take $W\to\infty$. Due to the factorization (or clustering) property of pure states 
\cite{georgii_book}, the position-averaged product of correlations $\langle s_i 
s_j\rangle_{\Gamma_{{\mathcal J}\varepsilon}} \langle s_i s_j\rangle_{\Gamma_{{\mathcal J}\varepsilon'}}$ for two 
pure states $\varepsilon$, $\varepsilon'$ tends to the square of the overlap,
\begin{equation}
q_{\varepsilon\varepsilon'}=\lim_{W\to\infty}\frac{1}{W^d}\sum_{{\bf r}_i\in\Lambda_W}\langle s_i 
\rangle_{\Gamma_{{\mathcal J}
\varepsilon}}\langle s_i \rangle_{\Gamma_{{\mathcal J}\varepsilon'}}\;,
\end{equation}
of the pure states, and by translation invariance of the joint distribution of $({\mathcal J},\varepsilon,\varepsilon')$ and 
the ergodic theorem for translation averages, the limit exists and is translation invariant. Then 
\begin{equation}
\lim_{r\to\infty}\lim_{t\to\infty} C_4(r,t)=\left[ \int d\varepsilon \,\int d\varepsilon' \,\mu_{\mathcal J}^{\rm M}(\varepsilon) 
\mu_{\mathcal J}^{\rm M}(\varepsilon')\,q_{\varepsilon\varepsilon'}^2   \right]_{\nu({\mathcal J})},
\end{equation}
and if there is at least one $\delta$-function in $\mu_{\mathcal J}^{\rm M}$, putting nonzero weight on one pure state, say 
$\varepsilon_0$, (and another for its global spin flip), the non-vanishing of the self-overlap $q_{\varepsilon_0\varepsilon_0}$
of that pure state when $T<T_c$ implies that $\lim_{t\to\infty}C_4(r,t)$ tends to a nonzero constant as $r\to\infty$, which 
is all we needed to show. If there are no such $\delta$-functions, then the limit will be zero if the overlaps of distinct 
pure states drawn independently from $\mu_{\mathcal J}^{\rm M}$ are almost always zero. That is what occurs under 
the hypothesis in Theorem 2 of NS \cite{Newman1999}, and in that case $\kappa_{\mathcal J}^{\rm M}=
\mu_{\mathcal J}^{\rm M}$. (It is also what occurs for $C$ in the 
equilibrium case in the RSB theory \cite{Read2014} where, as we mentioned already, $C(r)\to q(0)^2=0$, while it is believed
that the spin-glass correlation function in eq.\ (\ref{q2}) tends to $q^{(2)}>0$ in RSB because the pure-state decomposition 
of each Gibbs state $\Gamma_{\mathcal J}$ is countable.) It might appear 
that the statements about the pure-state structure of the MMAS could depend on $\mathcal J$, however, because of 
{\em translation} ergodicity of $\nu({\mathcal J})$, the total weight of the $\delta$-functions is the same for almost every 
$\mathcal J$, and so the character of the pure-state structure is the same for 
$\nu({\mathcal J})$-almost every $\mathcal J$. What we term trivial pure-state structure of the MMAS is of course 
not necessarily a completely trivial pure-state decomposition of the MMAS, but our use of the term is the most natural 
one for the behavior of the MMAS correlation function, and includes the SD case. 

\subsection{Statics-dynamics relation}

It was proposed in Refs.~\cite{Manssen2015-2, Wittmann2016} that the trivial or nontrivial nature of the pure state 
structure can be probed with Monte Carlo dynamics using the dynamically generated MMAS correlation function
in Eq.\ \eqref{eq:C4_equation}. The system is evolved in time with Monte Carlo 
dynamics: for a given timestep $t \rightarrow t + 1$, each lattice site of the total $N$ sites is visited, and a Metropolis 
accept/reject move is made for a single spin flip proposal according to the finite-size Gibbs distribution Eq.\ (\ref{finsizgibbs}). 
Based on the relations between static and maturation metastate averages already discussed, 
Ref.~\cite{Wittmann2016} conjectured a \emph{statics-dynamics} relation (see also Ref.\ \cite{White2006})
\begin{equation}
\alpha_s=\alpha_d,
\end{equation} 
and found empirically that, in the one-dimensional model with $\sigma=0.625$ (note this is less than $2/3$),  
$\alpha_d$ is in quantitative agreement with the value $\alpha_s=1/2$ that would be expected on the basis of the preceding
statements and conjecture. While it is unknown whether this conjectured equality always holds, $\alpha_s$ (and 
$\alpha_d$) will be zero for trivial pure state structure of the (M)MAS, and is expected to be nonzero for nontrivial. 

\subsection{Other recent work}

In a recent paper~\cite{Moore2021}, it was suggested that the results for $C_4$ may be affected by a crossover 
from RSB-like to  SD behavior as $r$ and $t$ increase. It was further suggested~\cite{Moore2021}, 
with reference also to~\cite{Baity-Jesi2017}, that when the SD picture is correct for equilibrium, and so applies 
at $r\ll\xi(t)$, the value of the exponent is given by $\alpha_d=2\theta$, where 
$\theta$ is the stiffness exponent of SD theory (in the notation of Ref.\ \cite{Fisher1988}). This value was 
obtained~\cite{Moore2021} from a calculation within SD theory of the decay exponent of a {\em connected} 
(or truncated) correlation function that describes the non-linear susceptibility and which in SD theory decays 
to zero as $r\to\infty$ \cite{Fisher1988}. However, the equilibrium correlation function $[\langle s_i 
s_j\rangle^2]_{\nu({\mathcal J})}$ (for trivial metastate) should tend to $q^{(2)}$ as $r_{ij}\to\infty$
(with power-law correction at finite $r$), where $q^{(2)}$ is not small when $T$ is well below $T_c$, and it is not clear 
in Ref.\ \cite{Moore2021} why this constant has been dropped; hence we do not believe 
that this argument establishes a relation between $\alpha_d$ (or $\alpha_s$) and $\theta$.


\section{Simulation results}
\label{sec:results}

\begin{table}
\caption{Exponents $\alpha_d$ and dynamic exponents z for the best fit collapse results of Fig.~\ref{fig:alpha_sig}. We also list 
the lattice size $N$ used in the collapse analysis and the maximum time $t_{max}$ reached in the simulations. $r_{min}$ values 
are listed as used for the best fit analysis. The fitting procedure is discussed in detail in Appendix~\ref{app:best_fit}.}
\centering
\begin{tabular}{c c c c c c c c c c c c c c c}
\hline\hline
$\sigma$  & $\alpha_d$ &$\delta \alpha_d$ & z & $\delta \textrm{z}$ & $T$ & $N$ & $r_{min}$ & $t_{max}$& $\xi(t_{max})$ 
\\ [0.9ex] 
\hline
0.610 & 0.571 &0.026&1.095&0.145 & 0.740 & $2^{26}$ &$2^{8}$& $2^{14}$ &3210\\
0.625&0.501&0.009&1.439 &0.052& 0.740 & $2^{26}$&$2^{7}$& $2^{14}$& 1415\\
0.667&0.418 &0.009&2.489&0.055 & 0.500 & $2^{24}$ &$2^{6}$& $2^{17}$ & 335\\
0.685 &0.367&0.009&2.529&0.047& 0.544 & $2^{24}$ &$2^{6}$& $2^{16}$ & 290\\
0.720 &0.327&0.011&2.739&0.057& 0.544 & $2^{22}$ &$2^{6}$& $2^{17}$ & 210\\
0.784 &0.202&0.013&3.464&0.102& 0.544 & $2^{18}$ &$2^{5}$& $2^{20}$ & 150\\
0.840&0.157&0.011&4.413&0.082& 0.450 & $2^{18}$ &$2^{4}$& $2^{20}$ &65 \\
0.896&0.127&0.017&5.316& 0.120& 0.400 & $2^{18}$ &$2^{4}$& $2^{20}$ &35\\[0.9ex]
\hline
\end{tabular}
\label{table:Values}
\end{table}

Our simulations are performed with a one-dimensional model introduced in Ref.~\cite{Leuzzi2008} where, on average, a site 
$i$ only has $z_{b}$ neighbors. The basic idea is to use diluted bonds such that, for an edge $\{i,j\}$, the bond $J_{ij}$ 
is nonzero (or the edge is occupied) with probability $P_{i,j}\propto 1/R_{ij}^{2\sigma}$ where $R_{ij}=(N/\pi)
\sin(\pi|i-j|/N)$ is the chord distance between sites $i$ and $j$ (whereas the lattice distance is $r_{ij}=|i-j|$ for $|i-j|<N/2$ 
and $r_{ij}=N-|i-j|$ otherwise), and the occupation numbers, which are $0$ or $1$ for each edge, are independent. The coefficient 
in $P_{i,j}\propto 1/R_{ij}^{2\sigma}$ is chosen so that the expected number of occupied edges is $Nz_b/2$ ($P_{ij}>1$ can be avoided
by softening the dependence on $R_{ij}$ at short distance, but keeping the asymptotic form at large $R_{ij}$). Given the set 
of occupied edges, those edges are then assigned values $J_{ij}$ independently from a Gaussian distribution with mean zero and 
variance unity (thus they are indeed nonzero with probability one), while unoccupied edges have $J_{ij}=0$. Note that the bonds 
$J_{ij}$ then satisfy eq.\ (\ref{kotdef}) for all $i$, $j$, and are independent random variables, as in the model of Ref.\ 
\cite{Kotliar1983}. In practice, as in Refs.\ \cite{Leuzzi2008} and \cite{Wittmann2016}, we will use a modified 
definition that is much less costly to implement at large lattice sizes. The interactions $\mathcal{J}=\left(J_{ij}\right)_{\{ij\}}$ for a given 
disorder realization, lattice size $N$, and coordination number $z_{b}$ are determined with the following procedure: (i) A site $i$ 
is chosen uniformly at random from the $N$ lattice sites. (ii) A site $j$ is then selected with probability 
$\widetilde{P}_{i,j}\propto 1/R_{ij}^{2\sigma}$, where now $\sum_j\widetilde{P}_{i,j}=1$ for the given lattice size $N$. 
(iii) If the edge $\{i,j\}$ 
does not already have a nonzero bond then we select one for this edge independently with a Gaussian distribution with mean zero 
and variance unity. If the edge already has a nonzero bond then we return to step (i) without modifying that bond. (iv) This process 
is repeated until there are $N z_b/2$ nonzero bonds. In the resulting model, the bonds again have mean zero and variance as in 
eq.\ (\ref{kotdef}) asymptotically, as stated there, and are uncorrelated but not strictly independent, because the occupation probabilities 
are no longer independent; in particular, the number of occupied edges is fixed, not random. We made comparisons of the results from 
the two models for some parameter values, and found that the differences were very small.

\begin{figure}
\begin{center}
\includegraphics[scale=0.75]{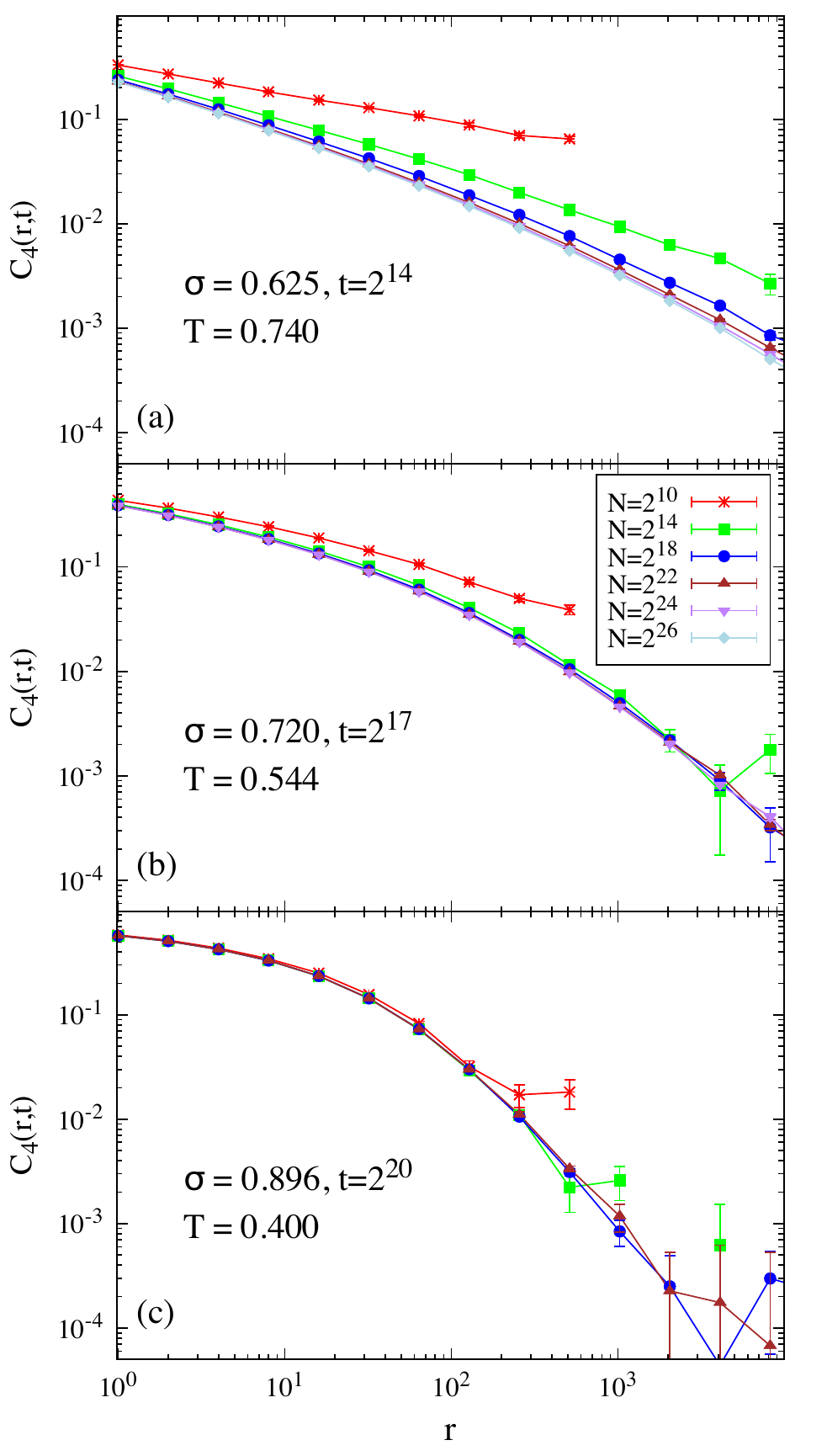}
\end{center}
\caption{$C_{4}(r,t)$ as a function of distance along the lattice $r$ for multiple lattice sizes $N$ for the largest times reached in 
the simulations for a few representative interaction parameters (a) $\sigma=0.625$, 
(b) $\sigma=0.720$, and (c) $\sigma=0.896$. We see that finite-size effects are well controlled already at 
$N=2^{14}$ for the largest $\sigma$ values but we require lattice sizes $N=2^{26}$ for $\sigma=0.625$.}   
\label{fig:lattime}
\end{figure}

The simulations were performed with coordination number $z_b=6$ for eight power-law interaction exponents ranging from 
$\sigma=0.61$ to $\sigma=0.896$. These simulations 
reach times of at least $t=2^{14}$ in all cases and $t=2^{20}$ for some interactions and lattice sizes. For each lattice size 
and coupling constant, we used between $N_s=80$ and $N_s=2000$ disorder
realization samples with $N_r=2$ real replicas for each realization
initialized for $t=0$ with spin configurations drawn independently and
randomly.

\begin{figure*}
\includegraphics[scale=0.75]{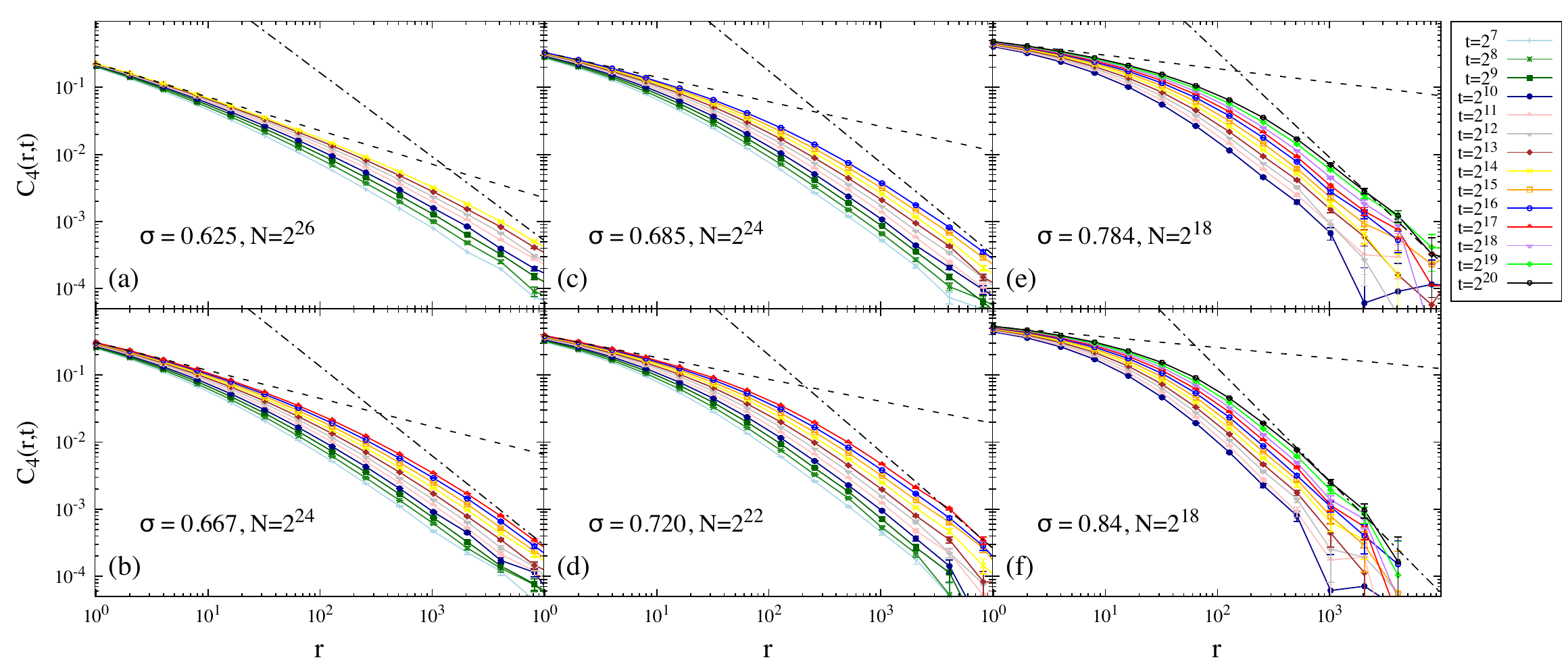}
\caption{$C_{4}(r,t)$ as a function of distance along the lattice $r$ for a given lattice size (with controlled finite-size effects) 
for varying times. Panel (a) shows interaction $\sigma=0.625$ for lattice size 
$N=2^{26}$ up to time $t=2^{14}$. Panel (b) shows results at $\sigma=2/3$ for lattice size $N=2^{24}$ up to time 
$t=2^{17}$. Panels (c)-(f) are for interactions in the regime $\sigma>2/3$, which is above the upper critical range. We 
also show the short-distance power-law behavior for each interaction $\sigma$ (dashed lines) and the large-distance 
power-law behavior (dashed-dotted lines) for the largest available value of $t$.} 
\label{fig:compare_time}
\end{figure*}

\begin{figure*}
\begin{center}
\includegraphics[scale=0.76]{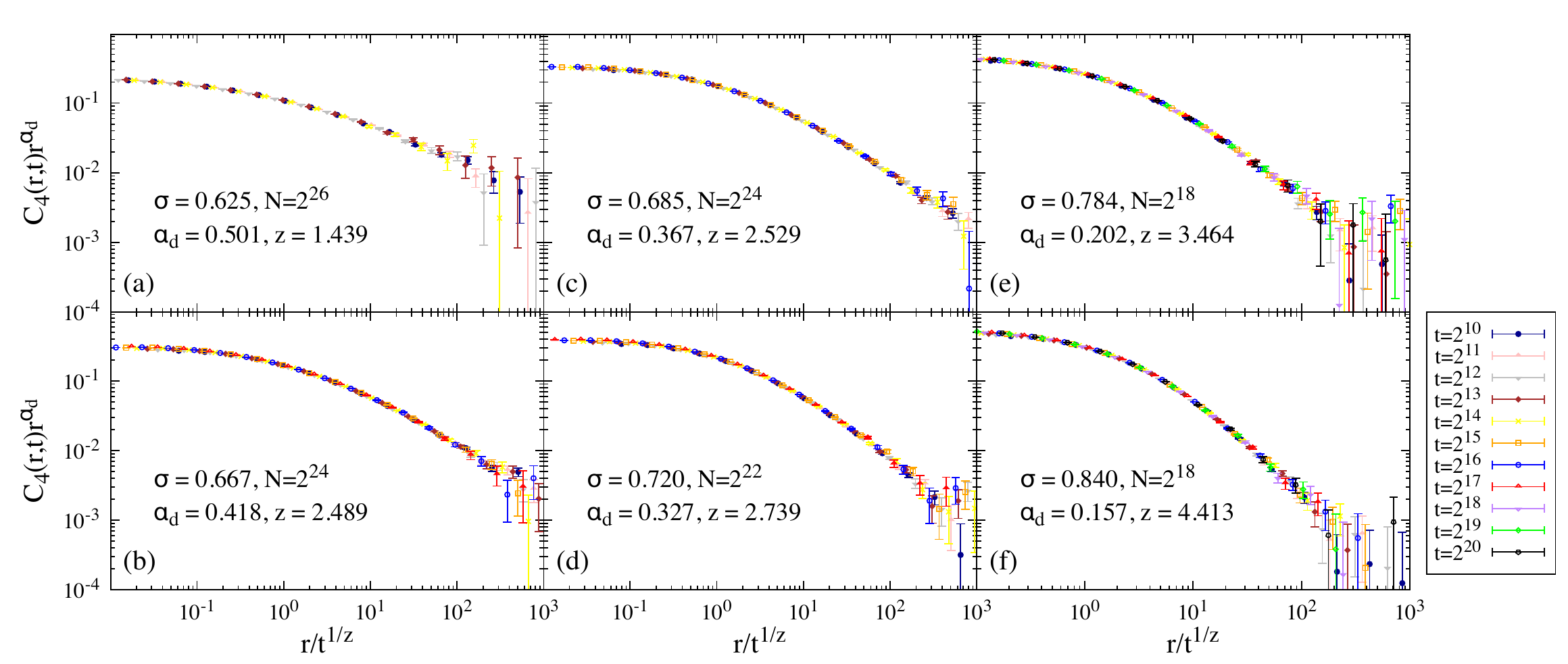}
\end{center}
\caption{Data collapse results for multiple interaction parameters from (a) $\sigma=0.625$, below the upper critical range, 
to (f) $\sigma=0.840$, well above the upper critical range. The collapse is performed with the ansatz 
$C_{4}(r,t)=\frac{1}{r^{\alpha_d}}f\left (\frac{r}{\xi(t)} \right)$ for the listed lattice sizes where $\xi(t)\propto t^{1/z}$. The best fit values used in the collapse are 
determined as discussed in App.~\ref{app:best_fit}.}   
\label{fig:collapse}
\end{figure*}

The temperatures used in the simulations were well below $T_c$. For the largest interaction exponent, $\sigma=0.896$, 
simulations were performed for $T=0.400$ given the critical temperature $T_c\simeq0.795$ of Ref.~\cite{Larson2013} 
where finite-size scaling of the static spin-glass susceptibility was used to determine $T_c$. For $\sigma=0.784$ we used 
$T=0.544$ given $T_c\simeq 1.36$ of the same work, Ref.~\cite{Larson2013}. For $\sigma=0.625$, $T_c=1.85(2)$ from 
Ref.~\cite{Wittmann2016} and, following this work, we performed simulations for $T=0.740$. The expected monotonic 
increase in critical temperature with decreasing interaction exponent guided our choices for the remaining couplings simulated 
(see Table~\ref{table:Values} for each temperature simulated and a summary of the simulation parameters and findings).

\begin{figure*}
\begin{center}
\includegraphics[scale=0.5]{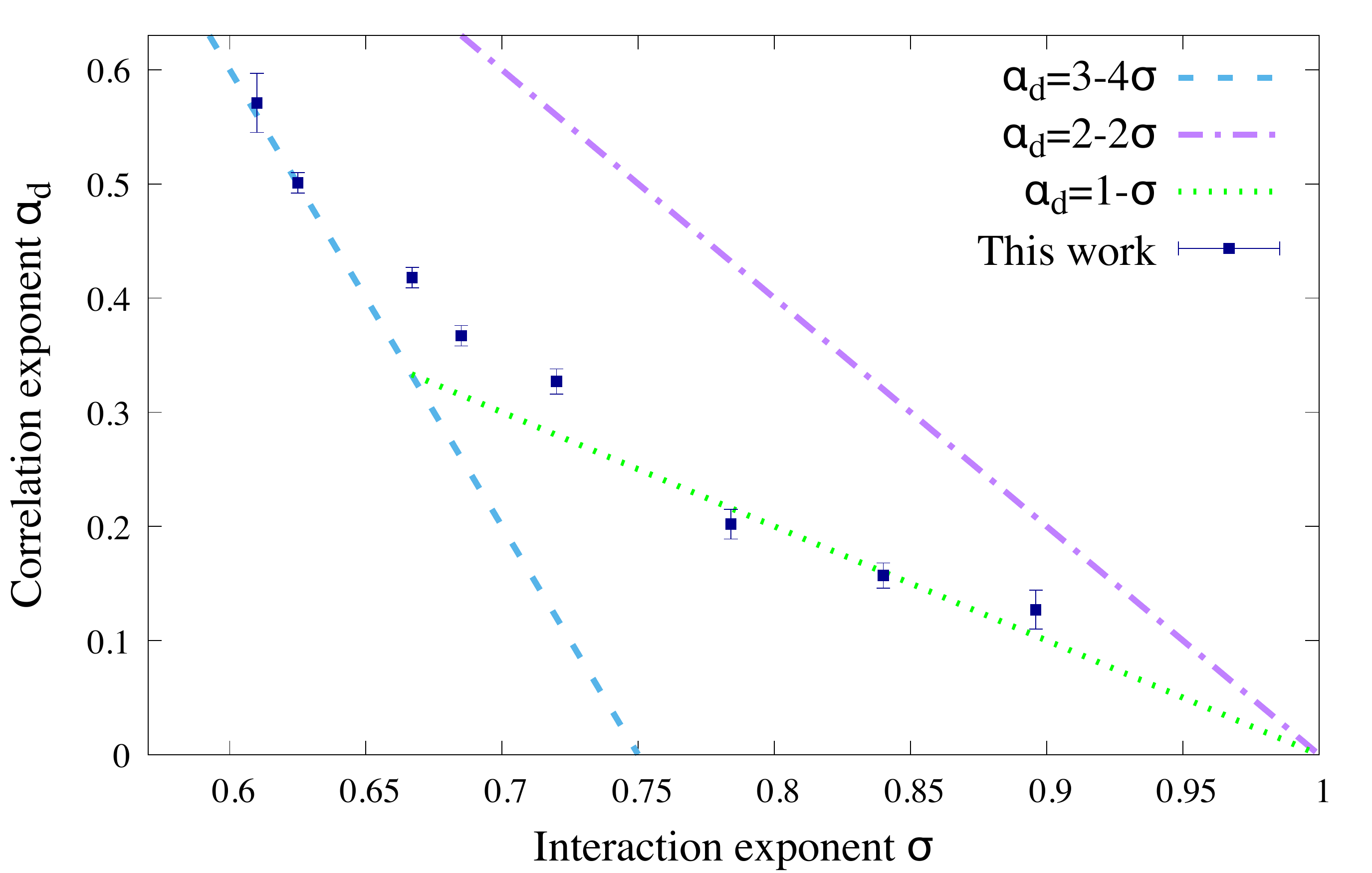}
\end{center}
\caption{Exponent $\alpha_d$ (dark blue squares) describing the decay of $C_{4}(r,t)$ as a function of interaction power-law 
exponent $\sigma$.  $\alpha_d>0$ for all interactions studied, suggesting a nontrivial metastate at low $T$ for $\sigma < 1$. 
We clearly see the predicted behavior $\alpha_d=3-4\sigma$ (light blue dashed line) for $\sigma < 2/3$. The dependence of 
$\alpha_d$ on $\sigma$ approaches $\alpha_d=1-\sigma$ (light green dotted line) for $\sigma > 2/3$. We also show 
the bound $\alpha_d \leq 2-2\sigma$~\cite{Read2021} (purple dashed dotted line) which is satisfied by the simulation results.}   
\label{fig:alpha_sig}
\end{figure*}

For an unbiased estimate of $C_{4}(r,t)$ of Eq.~(\ref{eq:C4_equation}) ($r \equiv r_{ij}$ by translation invariance) two replicas 
(an Ising spin is now denoted as $s_{i}^{\gamma}$ where $\gamma\in\left \{a,b\right \}$ labels the replica $a$ or $b$) of 
each disorder realization are independently simulated. Each replica has the same quenched couplings $\mathcal{J}$ but independent random initial spin configurations which are then independently evolved in time. For each of 
the $N_{s}$ disorder realization samples, we calculate the estimator
\begin{equation}\label{eq:practical_C4}
C_{4}(r,t)=\left [\frac{1}{N}\sum_{i=1}^{N} s_{i}^{a}(t)s_{i+r}^{a}(t)s_{i}^{b}(t)s_{i+r}^{b}(t) \right 
]_{\nu(\mathcal{J})}\;.
\end{equation}
In Fig.~\ref{fig:lattime} we show $C_{4}(r,t)$, calculated as in Eq.~\eqref{eq:practical_C4}, as a function of lattice size for three 
representative interactions $\sigma \in \left \{0.625,0.720,0.896\right \}$. For each coupling multiple lattice sizes $N\in \left \{ 
2^{10}, 2^{14}, 2^{18}, 2^{22}, 2^{24}, 2^{26} \right \}$ were simulated to investigate finite-size effects. We see from 
panel (c) of Fig.~\ref{fig:lattime} that finite-size effects are controlled for $\sigma=0.896$ already with lattice size $N=2^{14}$ 
for the time shown $t=2^{20}$ (results only up to lattice size $N=2^{22}$ were obtained for this time). For the smallest 
interaction exponent shown, $\sigma=0.625$, the finite-size effects are much more substantial even for $t=2^{14}$ requiring a 
lattice size $N=2^{26}$ as previously found in Ref.~\cite{Wittmann2016}. The reduction of finite-size effects with increasing 
$\sigma$ (for a given simulation time $t$), due to the faster power-law decay of the probability of the presence of a bond for 
$\{i,j\}$ with increasing $r$, has allowed for multiple $C_{4}(r,t)$ results with $\sigma > 2/3$. However, the larger $\sigma$ 
values require much larger times to achieve appreciable correlation lengths $\xi(t)$ and so similar computational effort was 
required for each interaction exponent $\sigma$.

With the finite-size effects controlled for each interaction exponent, the time dependence of $C_{4}(r,t)$ is analyzed. The results 
are plotted in Fig.~\ref{fig:compare_time}. The large-time and short-distance behavior for each interaction is fit with a power law 
(dashed line) in each panel for (a) $\sigma=0.625$ with lattice size $N=2^{26}$ through (f) $\sigma=0.84$ with lattice size 
$N=2^{18}$. The large-time and large-distance behavior is also fit by a power law (dashed-dotted line) showing the expected 
$C_{4}(r,t)\propto1/r^{2\sigma}$ dependence for the long-range model for $r \gg \xi(t)$ \cite{Wittmann2016}. We observe 
a crossover from short-range to long-range power-law dependence. The crossing of the two power-law curves for the 
maximum simulated time is used to give a rough estimate of the dynamic correlation length $\xi(t)$ reached (see 
Table~\ref{table:Values} for these estimates). From the raw data of Fig.~\ref{fig:compare_time} we perform a data 
collapse based on the scaling ansatz of Eq.~\eqref{eq:decay} with 
the product of $C_{4}(r,t)$ and $r^{\alpha_d}$ on the y-axis and $r/t^{1/z}$ on the x-axis and 
present the best fit collapse results in Fig.~\ref{fig:collapse}. The collapse for each value of $\sigma$ supports the scaling ansatz. 
We discuss our method for extracting the best fit collapse parameters and the associated errors in detail in 
Appendix~\ref{app:best_fit}.

The final results of the analysis are listed in Table~\ref{table:Values}. 
In Fig.~\ref{fig:alpha_sig} we show the best fit collapse values for $\alpha_d$ as a function of interaction power-law exponent 
$\sigma$. We show the prediction obtained using RSB theory together with the conjectured statics-dynamics relation, 
$\alpha_d=3-4\sigma$ \cite{Read2014,Wittmann2016} for $\sigma\leq 2/3$
(blue dashed line). We also show an upper bound $\alpha_d=2-2\sigma$ (purple dashed-dotted line); 
strictly speaking, this bound is obtained \cite{Read2021} from an upper bound $\alpha_s'\leq 2-2\sigma$ on the scaling, 
$\ln{\cal N}(W)\sim W^{\alpha_s'}$,
of the logarithm of the number $\cal N$ of pure states that could be seen in a window of size $W$ in any Gibbs state, 
and $\alpha_s'$ was conjectured to equal $\alpha_s$ \cite{Read2014}, as it does also in some of the models in Ref.\ 
\cite{White2006}. The same bound can also be obtained in another way: $\alpha=2-2\sigma$ is the decay exponent 
for the spin-glass correlation function at $T=T_c$ in this model in equilibrium \cite{Kotliar1983}, and one would expect 
a slower decay for the (M)MAS correlations below $T_c$. Finally, we show the line $\alpha_d=1-
\sigma$ (green dotted line), which interpolates between the expected value $1/3$ at $\sigma=2/3$ and $0$, which 
might be expected as $\sigma\to 1$ and which agrees with the upper bound. We see strong agreement of the best fit collapse 
results with $\alpha_d=3-4\sigma$ from the two values ($\sigma=0.625$ and $\sigma=0.61$) in the 
expected regime below the upper critical range $2/3$. This is in agreement with the previous study of Ref.~\cite{Wittmann2016} 
and supports both the statics-dynamics conjecture and the result from RSB  for $\sigma < 2/3$.  As $\sigma$ is increased 
beyond $\sigma=2/3$ we find that the best fit values for $\alpha_d$ remain nonvanishing and the line 
$\alpha_d =1-\sigma$ is a good fit to the three highest values of $\sigma$. This supports the theories with nontrivial 
MASs in the regime beyond the upper critical range $2/3$ also, and is in contrast to the expected result $\alpha_d=0$ 
of the SD picture. The data are not as close to either line where they intersect at $2/3$, and instead suggest 
a smooth curve. By analogy with critical phenomena, this might be due to
logarithmic corrections at the boundary value $\sigma = 2/3$.

The small statistical errors notwithstanding, our confidence in the results decreases as $\sigma$ approaches $1$, 
where we were only able to reach rather short correlation lengths, and $C_4$ does not decrease much as $r$ increases 
before $\xi(t)$ is reached, due to the small $\alpha_d$. This means that systematic errors could be much more significant 
as $\sigma\to1$ (e.g.\ because of corrections to scaling), and the results could change for larger $t$. It is difficult to say 
with confidence that we rule out the SD picture for $\sigma$ close to 1
unless we can see that $\lim_{t\to\infty}C_4(r,t)\ll q^{(2)}$ 
at sufficiently large $r$; this is not the case for $r<\xi(t)$ at our largest $\sigma$.

\section{Conclusion}
\label{sec:conclusion}

In this work we have extended the study of Ref.~\cite{Wittmann2016} for a 1D diluted long-range model to interaction
exponents $\sigma$ other than $\sigma=0.625$, considered in that work. We have performed dynamics simulations 
following a quench to temperatures below the critical temperature for interactions with range exponent $\sigma$ both below 
($\sigma<2/3$) and above ($\sigma>2/3$) the upper critical range. For all interactions considered we determined the best 
fit scaling exponents $z$ and $\alpha_d$. The best fit collapse value of $\alpha_d$ is found to be nonzero for all 
interactions considered indicating that the pure state structure is nontrivial (i.e.\ not the scaling-droplet picture) even above the 
upper critical range. We found evidence with multiple interaction exponents for the statics-dynamics equality conjecture 
$\alpha_d=\alpha_s$ ($\alpha_d=3-4\sigma$) for $\sigma<2/3$, which previously was quantitatively addressed only for 
$\sigma=0.625$ in Ref.~\cite{Wittmann2016}. Further, we found empirically that the correlation exponent approaches an 
interpolation curve $\alpha_d=1-\sigma$ as $\sigma\to1$.

It remains to determine in future studies if the statics-dynamics equivalence conjecture which is supported in this study 
can be strengthened or ruled out, including for the region $2/3<\sigma<1$. Both analytic and numerical approaches will 
be useful to address this. However, we emphasize again that both $\alpha_s$ and $\alpha_d$ are expected to be nonzero 
for nontrivial pure state structure and vanish for trivial. It will also be 
interesting to perform a similar numerical study in the presence of a magnetic field, where the phase diagram 
and the existence of an Almeida-Thouless (AT) line~\cite{Almeida1978} remains
uncertain~\cite{young:04,Larson2013,banosetal:12,baity-jesietal:14}.

\acknowledgments

The work of SJ was supported by the U.S. Department of Energy, Office of Science, Office of Basic Energy Sciences, 
Computational Materials Sciences program under Award Number DE-SC-0020177. NR acknowledges the support of 
NSF grant no.\ DMR-1724923. We also thank the Yale Center for Research Computing for the substantial computing time 
and resources necessary for the research presented here.

\begin{appendix}

\section{Best fit collapse}
\label{app:best_fit}
\begin{figure}
\begin{center}
\includegraphics[scale=0.61]{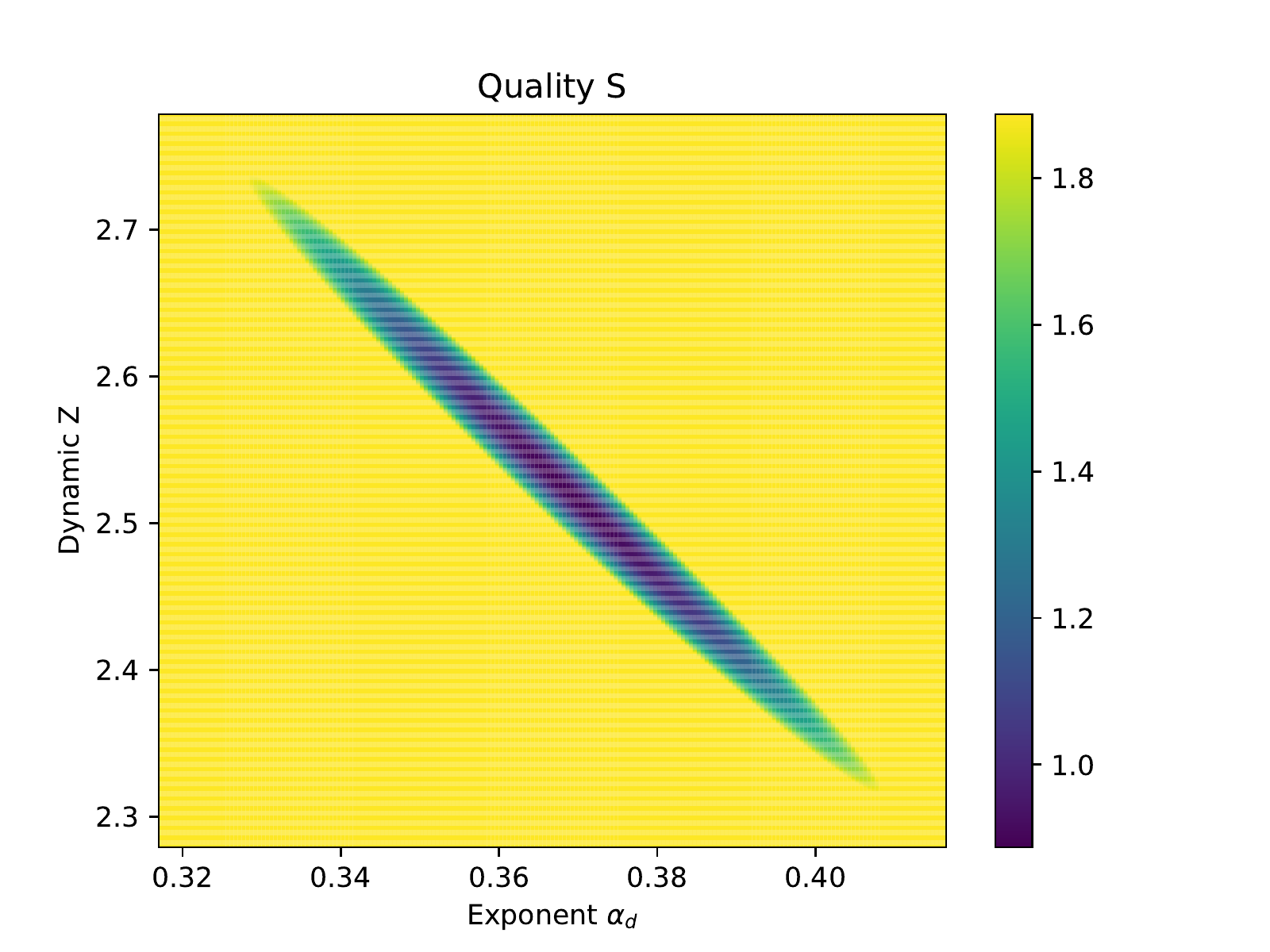}
\end{center}
\caption{Collapse quality $S$ heatmap for $\sigma=0.685$. The quality is considered good for $S\approx 1$ though, as 
discussed in the main text, it can be much smaller due to the highly correlated nature of the data. The quality shown was 
evaluated with data for $t_{min}=2^{13}$ to $t_{max}=2^{16}$ for $N=2^{24}$ and lattice distances $r_{min} = 2^{6}$.}   
\label{fig:heat}
\end{figure}

We do not know the scaling function $f(x)$ of Eq.~\ref{eq:decay} for the data collapse of Fig.~\ref{fig:collapse} of the main text 
but determine the best fit by introducing the quality $S$ quantitatively following the methods of Refs.~\cite{Kawashima1993, 
Houdayer2004}. For a fitting window of times $t\in(t_{min},t_{max})$ and lattice distances $r > r_{min}$, and for given 
collapse parameters $\alpha_d$ and $z$, we have an estimate for the collapsed correlation function, $C_{4}(r,t)r^{\alpha_d}$ 
denoted $y_{r,t}$ for each distance $r$ and time $t$ and a statistical error for this value $dy_{r,t}$. An estimate of the master 
curve function at $r/t^{1/z}$, denoted $Y_{r,t}$, can be made from the data $y_{r,t}$. For each point $y_{r,t}$ this
is done by fitting a
cubic polynomial to the three nearest scaled distances above, i.e.~$r'/t'^{1/z}>r/t^{1/z}$, and the three 
nearest scaled distances below, $r'/ t'^{1/z}<r/t^{1/z}$. The statistical error on the interpolated estimate is denoted 
$dY_{r,t}$. We define the collapse quality $S$ with
\begin{equation}\label{eq:quality}
S=\frac{1}{N_{L}}\sum_{r,t}\frac{(y_{r,t}-Y_{r,t})^{2}}{dy_{r,t}^2+dY_{r,t}^2}\;,
\end{equation}
where $N_{L}$ is the number of terms in the sum fixed by the size of the fitting window. Larger values of the quality indicate 
poor fits while the fit is considered good for $S\approx 1$. We show the quality $S$ as a function of $\alpha_d$ and $z$ for $
\sigma=0.685$ in Fig.~\ref{fig:heat} with $t_{min}=2^{13}$ and $t_{max}=2^{16}$ with $r_{min}=2^{6}$. 

\begin{figure*}
\includegraphics[scale=0.35]{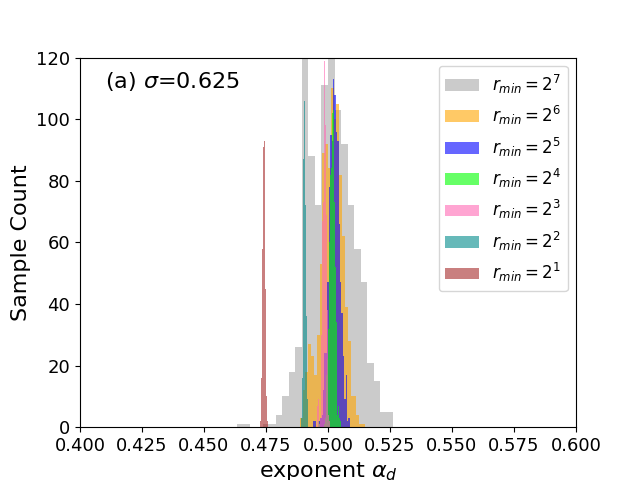}
\includegraphics[scale=0.35]{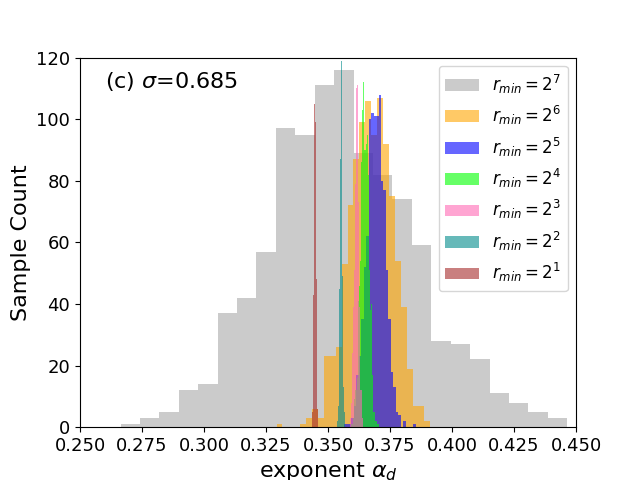}
\includegraphics[scale=0.35]{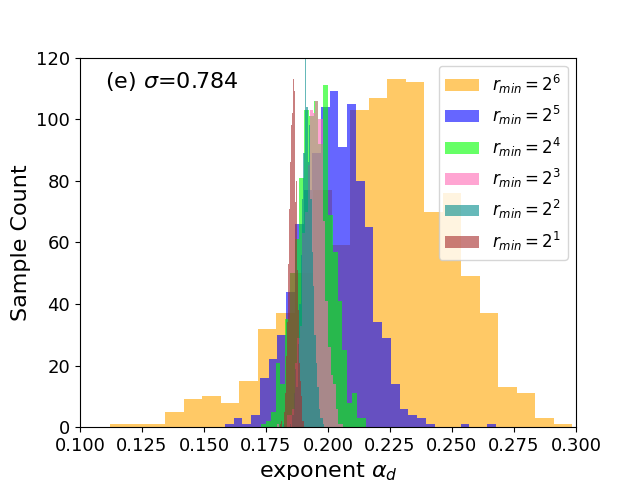}
\includegraphics[scale=0.35]{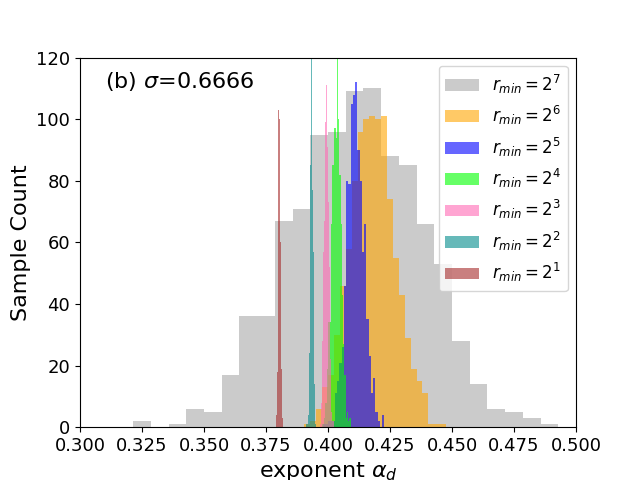}
\includegraphics[scale=0.35]{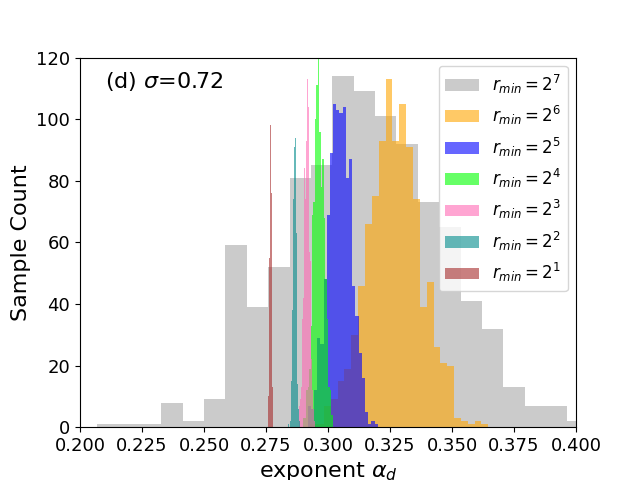}
\includegraphics[scale=0.35]{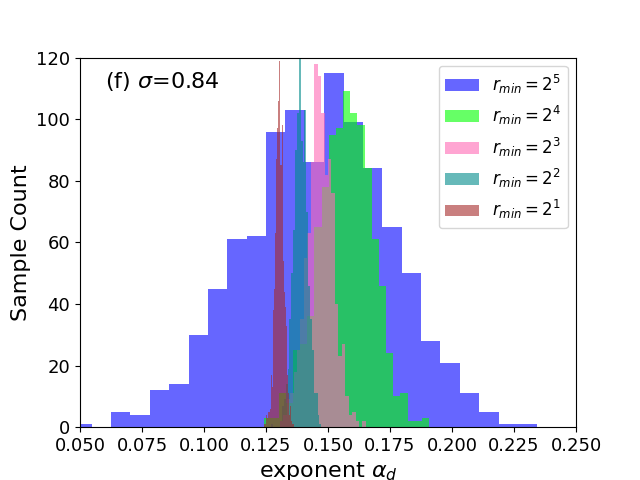}
\caption{Histograms for the best fit collapse exponent $\alpha_d$ with $N_{b}=1000$ bootstrap samples with interaction 
exponents (a) $\sigma=0.625$, (b) $\sigma=0.667$, (c) $\sigma=0.685$, (d) $\sigma=0.720$, (e) $\sigma=0.784$, and (f) 
$\sigma=0.840$. Each panel shows the bootstrap histogram for varying $r_{min}$. We see that the best estimate value for 
$\alpha_d$ varies significantly for small $r_{min}$, e.g., for $\sigma=0.625$ we observe variation up to $r_{min}\approx 
2^{4}$. The best fit collapse value of $\alpha_d=0.501$ (used in Fig.~\ref{fig:collapse}) and the error $\delta \alpha_d=0.009$ 
are determined with $r_{min}=2^{7}$. The values for each interaction exponent are given in Table~\ref{table:Values}.}   
\label{fig:Bootstrap}
\end{figure*}

The data considered here is highly correlated so the quality $S$ can be much smaller, e.g., the best fit $S_{min}\approx 0.1$ for 
$\sigma = 0.896$. The best fit quality $S_{min}$ can also be significantly larger if the interpolation is not sufficiently accurate 
as the statistical errors are relatively small on the correlation function $C_{4}(r,t)$ (this is why we use the cubic polynomial instead of 
the straight-line fit interpolation in Ref.~\cite{Houdayer2004}).
To determine a final estimate and error on the estimates for both $z$ and 
$\alpha_d$ with a given fitting window, specified by the parameters ($r_{min}$, $t_{min}$, $t_{max}$), we perform a 
bootstrap analysis with $N_{b}=1000$ bootstrap samples~\cite{Young2012}. For each bootstrap sample drawn from the 
underlying $N_s$ disorder realizations data, we determine the parameters $z$ and $\alpha_d$ which give the smallest fit value 
$S_{min}$ of Eq.~\eqref{eq:quality}. 

We show the bootstrap sample distributions in Fig.~\ref{fig:Bootstrap} which gave the best fit values and error bars for the parameters used in 
the collapse of Fig.~\ref{fig:collapse} of the main text. This was done for each interaction exponent $\sigma$ with $t_{max}$ as 
the largest time reached for the lattice size $N$ and multiple values for $r_{min}$. $t_{min}$ was chosen to be 
$t_{max}/2^{3}$ for each interaction exponent. We found the estimates to vary substantially for small $r_{min}$.
As an example consider $\sigma=0.685$ (top middle panel).
The 
best estimate value for $\alpha$ for this value of $\sigma$ varies for small $r_{min}$ up to $r_{min}\approx 
2^{4}$. The value of $r_{min}$ used for the final collapse parameters was selected by the requirement that $S_{min}$ saturates as 
a function of $r_{min}$. The best fit collapse value of $\alpha_d=0.367$ (used in Fig.~\ref{fig:collapse}) and the error $\delta 
\alpha_d=0.009$ were determined with $r_{min}=2^{6}$. For each value of the
interaction exponent,
the final collapse 
parameters found, and the value of $r_{min}$ used for the final estimates, are given in Table~\ref{table:Values} 
of the main text. \\

\end{appendix}

%

\end{document}